\documentclass[12pt]{article}
\usepackage{amsfonts}
\usepackage{amssymb}
\usepackage{graphics}
\usepackage[dvips]{graphicx}
\usepackage[]{psfig}
\usepackage{mathrsfs}
\usepackage{verbatim}
\usepackage{float}
\usepackage[english]{babel}
\def\simlt{\stackrel{<}{{}_\sim}}
\def\simgt{\stackrel{>}{{}_\sim}}

\title{
\vspace*{-1.3cm}
\begin{flushright}
\normalsize{
ANL-HEP-PR-06-55\\
EFI-06-10\\
}
\end{flushright}
\vspace{0.5cm}
\Large
\textbf{Soft Leptogenesis in Warped Extra Dimensions}
\vspace*{0.5cm}
\author{\textbf{Anibal D. Medina$^{a,b}$ and Carlos E. M. Wagner$^{b,c}$}\\
\\[0.5cm]
$^a$\normalsize\emph{Department of Astronomy and Astrophysics, The University of Chicago,}\\
\normalsize\emph{ 5640 S. Ellis Ave., Chicago, IL 60637, USA} \\
$^b$\normalsize\emph{HEP Division, Argonne National Laboratory,
9700 Cass Ave.,
Argonne, IL 60439, USA} \\
$^c$\normalsize\emph{Enrico Fermi Institute and Kavli Institute
for
Cosmological Physics,}\\
\normalsize\emph{The University of Chicago, 5640 S. Ellis Ave., Chicago, IL 60637, USA}}}

\date{}
\begin{document}
\setcounter{page}{0}
\maketitle
\vspace{0.5cm}
%
\begin{abstract}
We implement soft leptogenesis in a warped five dimensional
scenario with two branes on the orbifold boundaries coming from an
$S^{1}/Z_{2}$ symmetry, and supersymmetry broken on the IR brane.
The SM hypermultiplet fields (fermions and Higgs) live in the UV
brane and we allow the vector supermultiplets corresponding to the
gauge bosons and a hypermultiplet corresponding to the right
handed neutrino to live in the bulk. We assume that there are
Majorana mass terms for the right handed neutrino superfield fixed
on each brane and that there is a Yukawa term involving the right
handed neutrino, the left handed neutrino and the Higgs fixed on
the UV brane. Supersymmetry is broken by a constant
``superpotential'' on the IR brane, which induces an F-term for
the radion hypermultiplet. This F-term leads to a B-term for the
right handed sneutrinos as well as a soft SUSY breaking gaugino
mass in the 4D theory for the zero modes. The gaugino mass
naturally induces an A-term for the right handed sneutrino, left
handed sneutrino and the Higgs to be formed through gaugino
mediation with a non-trivial CP violating phase. Moreover, we show
that within the context of extra dimensions, the condition of
out-of-equilibrium decay and the phenomenological constraints on
the neutrino mass are both satisfied in a natural way, for UV
Majorana masses of the order of the fundamental scale of the
theory. Thus all necessary elements for soft leptogenesis are at
hand and we are able to predict a correct value for the baryon
asymmetry.
\end{abstract}
\thispagestyle{empty}

\newpage

\setcounter{page}{1}

\section{Introduction}

The idea of ``soft leptogenesis'' \cite{Grossman:2003jv},
\cite{D'Ambrosio:2003wy} which can explain the baryon asymmetry in
the universe is very attractive because of its simplicity. The
soft parameters provide the source of CP violation, not relying on
flavor physics like regular leptogenesis does. The presence of
these terms allows oscillations between right handed sneutrinos
and anti-sneutrinos which induce significant CP violation in
sneutrino-decay processes. In \cite{D'Ambrosio:2003wy},
\cite{Grossman:2004dz} a study of the parameter space was done,
but no  compelling model was addressed which could explain the
values obtained. A study of this sort was made in
\cite{Grossman:2005yi}, where gauge mediation and SUGRA effects
were used to explain the parameters. Our idea is to extend these
results to supersymmetric  warped extra dimensional theories with
one additional spatial extra dimension as in RS1
\cite{Randall:1999ee} \cite{Gherghetta:2000qt}. Working in a
supersymmetric scenario grant us the opportunity to change the
size of the extra dimension without worrying about the hierarchy
problem which is solved by supersymmetry.  Part of the motivation
for such an extension of the RS1 scenario comes from the point of
view of string theory where naturally supersymmetry and compact
extra dimensions are related, even though still there hasn't been
found any connection between this model and string theory.
Furthermore, not many models of leptogenesis in extra dimensions
have been worked out in the literature \cite{Pilaftsis:1999jk}.\\

In our framework, the Standard Model fermions and  Higgs
superfields live in what we  call the UV brane, the unwarped
brane. The right handed sterile neutrino and the gauge superfields
are in the bulk of the extra dimension and the radion superfield
acquires a non-vanishing F-term on the IR brane (warped brane)
which breaks supersymmetry.  In this context, the location of the
fields in the fifth dimension  provides a natural way to explain
the lepton asymmetry as well as to satisfy the constraints
necessary for leptogenesis to succeed. Further constraints from
neutrino masses and gravitino relic energy densities are satisfied
too. This leads to a strong predictive model which adjusts fairly
well to all constraints required for specific locations of the
fields in the extra dimension. The coupling constants and masses
have natural values except for the 5D right handed neutrino mass
parameter $M_{2}$ which must be some orders of magnitude smaller
than the GUT scale. One of the attractive points of this scenario
is that the condition of out of equilibrium decay of the right
handed sneutrinos is automatically obtained for values of the mass
parameter $M_1$ fixed on the UV brane of the order of the GUT
scale. This, in turn, leads to a neutrino mass of the order of
$10^{-3}$ eV, consistent with phenomenological constraints. As was
shown in \cite{Pomarol:2000hp}, the GUT scale for our model
coincides with the normal 4D GUT scale. This is related to the
fact that  though bare masses can be high, the KK masses always
 start at the order of the compactification scale $ke^{-k\pi R}$;
something that doesn't happen in flat extra dimensions.\\

The paper is organized as follows: in section 2 we introduce the
model in 5D superspace and we show how the F-term of the radion
superfield induces a gaugino mass. We write the Lagrangian in term
of the component fields and we give an interpretation to the
$\delta^{2}$ terms we find. To simplify the presentation, we work in a
one-generation model, discarding flavor indices, and we
calculate the 4D effective soft supersymmetry breaking
terms we will be working with. In section 3 we show where specifically
CP violation comes from, calculate the lepton asymmetry and discuss
the different constraints
on the model. We arrive at the conclusions in section 4.

\section{Superfield action in a warped 5D space}

Let us consider a 5D theory where the extra dimension is warped.
The extra dimension which we will denote with the letter $y$ is
compactified on an orbifold $S^{1}/Z_{2}$ of radius R, with $0\leq
y\leq \pi$ the angular coordinate. The metric is given by
\begin{equation}
ds^{2}=e^{-2R\sigma}\eta_{\mu\nu}dx^{\mu}dx^{\nu}+R^{2}dy^{2}
\end{equation}
where  $\sigma=k|y|$ and $1/k$ is the curvature radius. This space
corresponds to a slice of AdS$_{5}$. We promote R to a superfield
which corresponds to the 4D chiral radion superfield T that,
together with R, it is known to contain the fifth-component of the
graviphoton $B_{5}$, the fifth-component of the right handed
gravitino $\Psi^{5}_{R}$ and a complex auxiliary field $F_{T}$.
Higher-dimensional supersymmetric theories contain 4D
supersymmetry and therefore it is always possible to write them
using $\mathcal{N}=1$ superfields. We will write T as
\begin{equation}
T=R+iB_{5}+\theta\Psi^{5}_{R}+\theta^{2}F_{T}
\end{equation}
and we will take $<B_{5}>=<\Psi^{5}_{R}>=0$.\\

The five dimensional action in superspace for a hypermultiplet
corresponding to the right handed neutrino is given by,

\begin{eqnarray}
& & \mathcal{S}_{5}=\int d^{5}x\left(\int d^{4}\theta \frac{1}{2}(T+T^{\dagger}) e^{-(T+T^{\dagger})\sigma}(N^{\dagger}N+N^{c}N^{c \dagger})\right.\nonumber \\
& &
\left.+\int d^{2}\theta e^{-3T\sigma}N^{c}\left[\partial_{5}-(\frac{3}{2}-c_{\nu_{R}})T\sigma'\right]N +h.c.\right.\nonumber \\
& & \left.+\frac{1}{2}\int d^{2}\theta e^{-3T\sigma}N^{c}(-M_{1}
T)N^{c}\delta(y)+h.c.
+\frac{1}{2}\int d^{2}\theta e^{-3T\sigma}N^{c}(-M_{2} T)N^{c}\delta(y-\pi)\right.\nonumber\\
& &
\left.+h.c.-\int d^{2}\theta e^{-3T\sigma}\lambda LN^{c}HT\delta(y)+h.c.\right)\label{action}
\end{eqnarray}
where $N^{c}$ is the right handed neutrino chiral $\mathcal{N}=1$
hypermultiplet which together with N forms the 5D off-shell right
handed neutrino hypermultiplet. $N^{c}$ is even and N odd under $S^{1}/Z_{2}$ respectively. H is the Higgs hypermultiplet of
the up-type, L is the left handed neutrino hypermultiplet, M1 and
M2 are the Majorana masses in 5D, $\lambda$ is the Yukawa constant
in 5D and we parameterized the hypermultiplet mass as $c\sigma'$.
In our conventions, $d^{5}x=d^{4}xdy$. In components these fields
can be written as
\begin{eqnarray}
& & L=\tilde{\nu}+\sqrt{2}\nu\theta+F_{\nu}\theta^{2}\\
& &
N^{c}=\tilde{\nu}_{R}+\sqrt{2}\nu_{R}\theta+F_{N^{c}}\theta^{2}\\
& & H=H+\sqrt{2}h\theta+F_{H}\theta^{2}\\
& & N=\tilde{N}+\sqrt{2}\psi\theta+F_{N}\theta^{2}
\end{eqnarray}

The auxiliary field $F_{T}$ which  will be responsible for breaking
supersymmetry on the $y=\pi$ boundary, comes from the effective
Lagrangian \cite{Marti:2001iw}
\begin{equation}
\mathcal{L}_{4D}=-\frac{6M_{5}^{3}}{k}\int d^{4}\theta \phi^{\dagger}\phi(1-e^{-(T+T^{\dagger})k\pi})+\int d^{2}\theta\phi^{3}[W_{0}+e^{-3Tk\pi}W]+h.c.
\label{eq:compensator}
\end{equation}
where W and  $W_{0}$ are superpotentials at the orbifold positions
$y=0$ and $y=\pi$, $\phi$ is the compensator field and
$M_{P}^{2}=M_{5}^{3}(1-e^{-2k\pi R})/k$, with $M_{P}$ being the 4D Planck
mass. $W_{0}$ was introduced to cancel the cosmological constant
in the 4D theory,  and $|W_{0}|^{2}=e^{-4k\pi R}|W|^{2}$. This implies that
SUSY breaking is heavily suppressed on the Planck brane; therefore
we can assume that $F_{T}$ is localized on the IR brane and its
effective 4D form is
\begin{equation}
F_{T}=e^{-k \pi  R}\frac{W}{2\pi M_{5}^{3}}  .
\end{equation}

So far, we have not introduced the gauge field components. These,
however, play an essential role in the model analyzed in this
article, since the soft supersymmetry breaking parameters of the
Higgs and left-handed lepton chiral fields are generated via the
mechanism of gaugino mediation. The 5D vector superfield includes
two gauginos. One of these gauginos, which we will denote as
$\lambda_1$,  transforms in a vector supermultiplet together with
the gauge fields while the other, $\lambda_2$, forms the fermion
component of a scalar superfield transforming in the adjoint
representation of the group. The kinetic action for the vector
supermultiplet can be parameterized in terms of the fermion chiral
superfield $W^{\alpha} \simeq \lambda_1^{\alpha} + ...$, and has
the form
\begin{equation}
\mathcal{S}_{5}=\int d^{5}x\left[\frac{1}{4g_{5}^{2}}
\int d^{2}\theta TW^{\alpha}W_{\alpha}+h.c.+...\right]
\end{equation}

The radion F-term breaks SUSY inducing a localized gaugino mass given by
\begin{equation}
\mathcal{L}_{soft}=\frac{\delta(y-\pi)e^{-k\pi R}W\lambda_{1}\lambda_{1}}
{RM_{5}^{3}}+h.c.\label{gino}
\end{equation}
Redefining $\lambda_{i} \rightarrow e^{-2R\sigma}\lambda_{i}$, $i= 1,2$
to absorb the spin connection term, the equations of motion for the
gauginos are given by
\begin{eqnarray}
& &
ie^{R\sigma}\bar{\sigma}^{\mu}\partial_{\mu}\lambda_{2}+\frac{1}{R}(\partial_{5}+
\frac{1}{2}R\sigma')\bar{\lambda}_{1}=0 ,
\nonumber\\
& &
ie^{R\sigma}\bar{\sigma}^{\mu}\partial_{\mu}\lambda_{1}-\frac{1}{R}(\partial_{5}-
\frac{1}{2}R\sigma')\bar{\lambda}_{2}-\frac{W}{2M_{5}^{3}R}\delta(y-\pi)\bar{\lambda}_{1}=0  .
\label{gau2}
\end{eqnarray}
We solve these equations in the bulk, ignoring boundary effects
which will only play a role when imposing boundary conditions.
Looking for solutions of the form $\lambda_{i}(x,y)=\sum
\lambda^{(n)}f_{i}^{(n)}(y)$ and using the orthogonality condition
of the modes, Eq.~(\ref{gau2}) leads to the second order differential
equations\cite{Marti:2001iw}
\begin{equation}
\left[\frac{1}{R^{2}}e^{R\sigma}\partial_{5}(e^{-R\sigma}\partial_{5})-(\frac{1}{4}\pm\frac{1}{2})k^{2}\right]
f_{1,2}^{(n)}=e^{2R\sigma}m_{n}^{2}f_{1,2}^{(n)}
\end{equation}
with solutions
\begin{eqnarray}
& &
f_{1}^{(n)}(y)=\frac{e^{R\sigma/2}}{N_{n}}\left[J_{1}\left(\frac{m_{n}}{k}e^{R\sigma}\right)+b_{1}
(m_{n})Y_{1}\left(\frac{m_{n}}{k}e^{R\sigma}\right)\right]\\
& &
f_{2}^{(n)}(y)=\frac{\sigma'e^{R\sigma/2}}{kN_{n}}\left[J_{0}\left(\frac{m_{n}}{k}e^{R\sigma}\right)+
b_{2}(m_{n})Y_{0}\left(\frac{m_{n}}{k}e^{R\sigma}\right)\right]
\end{eqnarray}
where $b_{i}$ and $m_{n}$ will be determined by the boundary
conditions, and $N_{n}$ are normalization constants.\\

Taking into account the $Z_{2}$ assignment, $f_{i}^{(n)}$ must
fulfill the following conditions on the $y=0$ boundary
\begin{eqnarray}
& &
f_{2}^{(n)}|_{y=0}=0\\
& &
\left(\frac{d}{dy}+\frac{R}{2}\sigma'\right)f_{1}^{(n)}|_{y=0}=0
\end{eqnarray}
which imply $b_1(m_{n})=b_2(m_{n})=-J_{0}(m_{n}/k)/Y_{0}(m_{n}/k)$.
On the other hand, the presence of the Majorana gaugino mass on
the $y=\pi$ boundary in  Eq.~(\ref{gau2}) implies
\begin{equation}
f_{2}^{(n)}(\pi)= \frac{W}{4M_{5}^{3}}f_{1}^{(n)}(\pi)
\end{equation}
These conditions yield that the following determinant should vanish
\begin{equation}
 \det{\left( \matrix{
~~~J_{0}(x_{n})  &
Y_{0}(x_{n})\cr
J_{0}(x_{n}e^{k\pi R})-\frac{W}{4M_{5}^{3}}J_{1}(x_{n}e^{k\pi R}) &
~\,~Y_{0}(x_{n}e^{k\pi R})-\frac{W}{4M_{5}^{3}}Y_{1}(x_{n}e^{k\pi R}) \cr}
\right)} ,
\nonumber
\end{equation}
where $x_{n}=m_{n}/k$. From here we get the KK gaugino mass
spectrum. Solving this equation we find a non-zero value for the
zero mode gaugino mass. In the case that
$\eta\equiv\frac{W}{4M_{5}^{3}}\ll 1$ (small SUSY breaking) and
$x_{n}e^{k\pi R}\ll 1 $ (for the zero mode) we find
\begin{equation}
m_{\lambda_{1}}\approx -\frac{\eta}{\pi R}e^{-k\pi R}
\end{equation}
Multiplying numerator and denominator by $k$, we see that under
this conditions, the zero mode gaugino mass will be much smaller
than the KK mass scale, parameterized by $k e^{-k \pi R}$. It is
important to stress again that, contrary to the standard warped
extra dimension scenarios, the hierarchy is stabilized by
supersymmetry and therefore there is no need for the KK mode masses to
be close to the weak scale. In general the KK masses will be out
of the reach of the LHC. As we will show, the phenomenological
properties of this model will be similar to those of low energy
supersymmetry breaking with a light gravitino.\\

We are interested in obtaining the effective action for the right-handed
neutrinos. In order to do that, we need
to calculate the auxiliary field for $N^{c}$ and N.
From Eq.~(\ref{action}), we obtain,
\begin{eqnarray}
& &
F_{N^{c}}^{\dagger}=-\frac{e^{-R\sigma}}{R}\Bigg(\left[\partial_{5}-\left(\frac{3}{2}-
c_{\nu_{R}}\right)R\sigma'\right]\tilde{N}-M_{1}R\tilde{\nu}_{R}\delta(y)-
M_{2}R\tilde{\nu}_{R}\delta(y-\pi)\nonumber\\
& &
-\lambda\tilde{\nu}HR\delta(y)\Bigg)-\frac{1}{2R}\tilde{\nu}_{R}^{*}F_{T}(1-2R\sigma)\\
& &
F_{N}^{\dagger}=\frac{e^{-R\sigma}}{R}\left[\partial_{5}-\left(\frac{3}{2}+c_{\nu_{R}}\right)R\sigma'\right]
\tilde{\nu}_{R}-\frac{1}{2R}\tilde{N}^{*}F_{T}(1-2R\sigma)
\end{eqnarray}

Replacing these F-terms in Eq.~(\ref{action}) and  integrating over
superspace we get the following 5D Lagrangian for
$\tilde{\nu}_{R}$ and $\tilde{N}$
\begin{eqnarray}
& &
\mathcal{L}_{5D}=\sqrt{-g}(-|\partial_{M}\tilde{N}|^{2}-|\partial_{M}\tilde{\nu}_{R}|^{2}-
m_{N}^{2}\tilde{N}\tilde{N}^{*}-m_{N^{c}}^{2}\tilde{\nu}_{R}\tilde{\nu}_{R}^{*}\nonumber \\
& &
+\frac{e^{R\sigma}}{2R^{2}}\tilde{N}F_{T}(1-2R\sigma)(\partial_{5}\tilde{\nu}_{R}-(3/2+
c_{\nu_{R}})\sigma'R\tilde{\nu}_{R})+h.c-\frac{e^{R\sigma}}{2R^{2}}2(3/2-c_{\nu_{R}})
\sigma'RF_{T}\tilde{\nu}_{R}\tilde{N}\nonumber\\
& &
+h.c.-\frac{e^{R\sigma}}{2R^{2}}F_{T}\tilde{\nu}_{R}(1+4R\sigma)(\partial_{5}\tilde{N}
-(3/2-c_{\nu_{R}})\sigma'R\tilde{N})+h.c.-M_{1}^{2}\tilde{\nu}_{R}\tilde{\nu}_{R}^{*}\delta(y)^{2}
\nonumber\\
& &
-\lambda\lambda^{*}\tilde{\nu}\tilde{\nu}^{*}HH^{*}\delta(y)^{2}-M_{1}\tilde{\nu}_{R}^{*}
\lambda\tilde{\nu}H\delta(y)^{2}+h.c.\nonumber\\
& &
+\frac{e^{2R\sigma}}{2R}(2R\sigma-2)F_{T}F_{T}\sigma(\tilde{\nu}_{R}\tilde{\nu}_{R}^{*}+
\tilde{N}\tilde{N}^{*})\nonumber\\
& &
-M_{2}^{2}\tilde{\nu}_{R}\tilde{\nu}_{R}^{*}\delta(y-\pi)^{2}+\frac{\tilde{\nu}_{R}M_{1}}{R}
(\partial_{5}\tilde{N}^{*}-(3/2-c_{\nu_{R}})\sigma'R\tilde{N}^{*})\delta(y)+h.c.\nonumber\\
& &
+\frac{\tilde{\nu}_{R}M_{2}}{R}(\partial_{5}\tilde{N}^{*}-(3/2-c_{\nu_{R}})\sigma'R\tilde{N}^{*})
\delta(y-\pi)+h.c.+\nonumber\\
& &
\frac{\lambda\tilde{\nu}H}{R}(\partial_{5}\tilde{N}^{*}-(3/2-c_{\nu_{R}})\sigma'R\tilde{N}^{*})
\delta(y)+h.c.+\frac{M_{2}}{2}\sigma e^{\sigma
R}F_{T}\tilde{\nu}_{R}\tilde{\nu}_{R}\delta(y-\pi)+h.c.),
\label{lag}
\end{eqnarray}
where $m^{2}_{N,N^{c}}=(c_{\nu_{R}}^{2}\pm
c_{\nu_{R}}-15/4)k^{2}$. The delta-squared terms are similar to
those found by Horava in \cite{Horava:1996vs}. They are related to
the bulk-boundary coupling, and as pointed out in
\cite{Mirabelli:1997aj} they are necessary in order to have SUSY
conserved. They can also be thought as parameterizing the effects
induced by the sum over the KK towers. Let us show  this in the
explicit example of the relation between neutrino and sneutrino
masses.\\

The Majorana mass term for the right handed neutrino localized on the UV
brane is given by
\begin{equation}
\frac{1}{2}M_{1}\nu_{R}(x,0)\nu_{R}(x,0)=M_{1}\sum_{n,m}f^{(n)}_{R}(0)f^{(n)}_{R}(0)\nu_{R}^{(n)}(x)\nu_{R}^{(m)}(x)
\end{equation}
which can be interpreted in matrix form in the basis of
$(\nu_{R}^{(0)}(x),\nu_{R}^{(1)}(x),\nu_{R}^{(2)}(x), \dots)$ as
\begin{equation}
S= \left( \matrix{ ~~~ f^{(1)}_{R}(0)f^{(1)}_{R}(0) &
f^{(1)}_{R}(0)f^{(2)}_{R}(0) & f^{(1)}_{R}(0)f^{(3)}_{R}(0) &
\dots \cr f^{(2)}_{R}(0)f^{(1)}_{R}(0) &
f^{(2)}_{R}(0)f^{(2)}_{R}(0) & f^{(2)}_{R}(0)f^{(3)}_{R}(0)    &
\dots \cr ~\,~f^{(3)}_{R}(0)f^{(1)}_{R}(0) & f^{(3)}_{R}(0)
f^{(2)}_{R}(0) & f^{(3)}_{R}(0)f^{(3)}_{R}(0)    & \dots \cr
\vdots & \vdots & \vdots & \ddots \cr} \right)
\end{equation}
The same KK expansion can be done for the right handed sneutrino
with functions $g^{(n)}_{R}(y)$. Since the $g^{(n)}_{R}$ form a
complete orthonormal system we can expand the $\delta(0)$ in this
basis as
\begin{equation}
\delta(0)=\sum_{k}g^{(k)}_{R}(0)g^{(k)}_{R}(0)
\end{equation}
Therefore if we look at the SUSY mass term
\begin{eqnarray}
& & \int
dy\sqrt{-g}(M_{1}^{2}\tilde{\nu}_{R}\tilde{\nu}_{R}^{*}\delta(y)^{2})=\nonumber\\
& & \int
dy[\sqrt{-g}(M_{1}^{2}\tilde{\nu}_{R}\tilde{\nu}_{R}^{*}\delta(y))]\times
\delta(y)
\end{eqnarray}
After proper normalization, it will take the form
\begin{eqnarray}
& &
M_{1}^{2}\sum_{n,m}g^{(n)}_{R}(0)g^{(m)}_{R}(0)\delta(0)\tilde{\nu}_{R}^{(n)}(x)\tilde{\nu}_{R}^{(m)*}(x)=\nonumber\\
& &
M_{1}^{2}\sum_{n,m,k}g^{(n)}_{R}(0)g^{(k)}_{R}(0)g^{(k)}_{R}(0)g^{(m)}_{R}(0)\tilde{\nu}_{R}^{(n)}(x)\tilde{\nu}_{R}^{(m)*}(x)
\end{eqnarray}
which can be interpreted in the basis
 $(\tilde{\nu}_{R}^{(0)}(x),\tilde{\nu}_{R}^{(1)}(x),\tilde{\nu}_{R}^{(2)}(x), \dots)$ as
$M_{1}^2 S'\times S'$ with S' the mass matrix formed with the
$g^{(n)}(0)$ functions. We remind the reader that in AdS$_{5}$
background fields in the same supermultiplet must have different
masses which will lead them to have different dependence on the
fifth dimension. However, in the case of flat extra dimensions
$g^{(n)}_{R}(y)=f^{(n)}_{R}(y)$ and we see then that  this leads
to the conventional supersymmetric relations between the neutrino
and sneutrino mass matrices.
\\

\subsection{Sneutrino bilinear and trilinear SUSY breaking terms}

The mechanism of soft leptogenesis requires specific relations between
the soft supersymmetry breaking bilinear and trilinear terms of the
sneutrinos.
From the Lagrangian, Eq.~(\ref{lag}), we see that nor A-term, neither
B-term can be formed on the UV brane at tree-level. This has to
do with the fact
that  SUSY breaking is localized on the IR brane and that the
B-term is proportional to $\sigma$. However, since the right-handed
neutrino superfield propagates into the extra dimension, a
B-term will be naturally induced on the IR brane (we remind the reader that
the terms proportional to $F_{T}$ are localized on the IR brane).\\

 For the zero modes, Eq.~(\ref{lag}) reduces to
\begin{eqnarray}
& &
\mathcal{L}_{5D,0}=\sqrt{-g}(-|\partial_{M}\tilde{N}|^{2}-|\partial_{M}\tilde{\nu}_{R}|^{2}-M_{1}^{2}\tilde{\nu}_{R}\tilde{\nu}_{R}^{*}\delta(y)^{2}\nonumber\\
& &
-\lambda\lambda^{*}\tilde{\nu}\tilde{\nu}^{*}HH^{*}\delta(y)^{2}-M_{1}\tilde{\nu}_{R}^{*}\lambda\tilde{\nu}H\delta(y)^{2}+h.c.\nonumber\\
& &
+\frac{e^{2R\sigma}}{2R}(2R\sigma-2)F_{T}F_{T}\sigma(\tilde{\nu}_{R}\tilde{\nu}_{R}^{*}+\tilde{N}\tilde{N}^{*})-M_{2}^{2}\tilde{\nu}_{R}\tilde{\nu}_{R}^{*}\delta(y-\pi)^{2}+\nonumber\\
& & \frac{1}{2}M_{2}\sigma e^{\sigma
R}F_{T}\tilde{\nu}_{R}\tilde{\nu}_{R}\delta(y-\pi)+h.c.)\label{lag2}
\end{eqnarray}

We know that the right-handed massless zero mode for
$\tilde{\nu}_{R}$ satisfies the following equation
\begin{equation}
[\partial_{5}-(3/2+c_{\nu_{R}})T\sigma']g^{(0)}=0
\end{equation}
whose solution is $g^{(0)}=e^{(3/2+c_{\nu_{R}})T\sigma}/N_{0}$,
where $N_{0}$ is a normalization constant. Analyzing the form of the
kinetic term in Eq.~(\ref{lag2}), we can obtain the normalization factor for
the zero mode. Canonically normalizing the right handed sneutrino
field, we find that
\begin{equation}
\frac{1}{N_{0}^{2}}=\frac{2(1/2+c_{\nu_{R}})k}{e^{2(1/2+c_{\nu_{R}})k\pi
R}-1} .
\end{equation}
\\
Similarly, the normalization condition for the zero mode fields fixed on the
UV brane comes from
\begin{equation}
\int_{0}^{\pi}\frac{1}{N_{0}^{2}}\delta(y)Rdy=1
\end{equation}
 Therefore $N_{0}=\sqrt{R}$.\\

In order to derive the form of the $A$ and $B$ parameters of the sneutrino,
we need to determine the size of the effective Yukawa and Majorana masses
for the right-handed neutrino field.
If we look at the fermionic interactions for the superfields, we
obtain the following term
\begin{equation}
{\cal{L}}_{\rm Yukawa} \simeq
\lambda R(\nu\nu_{R}H+\nu_{R}h\tilde{\nu}+h\nu\tilde{\nu}_{R}) + h.c.
\end{equation}
After canonically normalizing, this term takes the form
\begin{equation}
{\cal{L}}_{\rm Yukawa} \simeq
\frac{\lambda
\sqrt{k (1 + 2 c_{\nu_{R}})}}{\sqrt{e^{2(1/2+c_{\nu_{R}})k\pi R}-1}}
(\nu\nu_{R}H+\nu_{R}h\tilde{\nu}+h\nu\tilde{\nu}_{R}) + h.c.
\end{equation}
Therefore, we identify the 4D Yukawa coupling constant
\begin{equation}
\lambda_{4}=\frac{\lambda
\sqrt{k(1 +2 c_{\nu_{R}})}}{\sqrt{e^{2(1/2+c_{\nu_{R}})k\pi
R}-1}}
\end{equation}
\\
We can do the same in the case of the Majorana mass. Then we get
for the fermionic part in the case of IR or UV Majorana term
\begin{eqnarray}
{\cal{L}}_M \simeq \frac{1}{2}M_{2}e^{2c_{\nu_{R}}k\pi
R}\nu_{R}\nu_{R} R \delta(y - \pi) +
\frac{1}{2}M_{1}R\nu_{R}\nu_{R} \delta(y)  .
\end{eqnarray}
Canonically normalizing these terms, we obtain the values of the
localized Majorana masses for the right-handed neutrino,
\begin{eqnarray}
& & M_{4,IR}=2\frac{(1/2+c_{\nu_{R}})kRe^{2c_{\nu_{R}}k\pi R}}
{e^{2(1/2+c_{\nu_{R}})k\pi R}-1}M_{2}  , \\
& &M_{4,UV}=2\frac{(1/2+c_{\nu_{R}})kR}{e^{2(1/2+c_{\nu_{R}})k\pi
R}-1}M_{1} .
\end{eqnarray}
We see that provided the right-handed neutrino zero mode is
localized towards the IR brane, $c_{\nu_R} > -1/2$, as we will
assume in our model, the localized Majorana mass in the
ultraviolet  will be much larger than the one in the infrared,
$M_{4,UV}\gg M_{4,IR}$. Thus the effective Majorana mass of the
right-handed neutrino, $M_4$, is dominated by the ultraviolet
term, $M_{4}\simeq M_{4,UV}$.\\

We are now prepared to compute the bilinear and trilinear soft
supersymmetry breaking terms in the effective four dimensional
effective theory. Replacing the zero-mode for $\tilde{N^{c}}$ in
Eq.~(\ref{lag}), integrating on the fifth dimension and
canonically normalizing we get the following 4D B-term (we define
the bilinear term in the soft Lagrangian as
$-\mathcal{L}_{soft,4D}=\ldots+\frac{1}{2}B_{4}M_{4}
\tilde{\nu}_{R}(x)\tilde{\nu}_{R}(x)+\ldots$, where $M_{4}$ is the
Majorana mass of the right handed neutrinos),
\begin{equation}
B_{4}=k\pi F_{T}\frac{M_{4,IR}}{M_{4,UV}} .
\end{equation}

\begin{figure}[h]
\centerline{
\psfig{figure=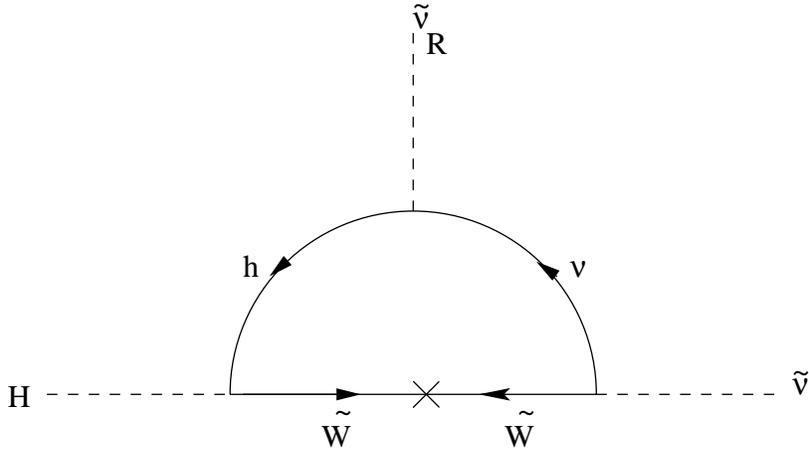,height=6cm,angle=0}}
\smallskip
\caption{ Feynman diagram for $A_{4}$ generation. \label{fig1} }
\end{figure}

On the other hand,  in the presence of phases in the gaugino mass
terms, massive gauginos will naturally induce an
$A_{4}$-term with a CP violating phase. We define the A-term as
$-\mathcal{L}_{soft,4D}=\ldots+A_{4}\lambda_{4}\tilde{\nu}_{R}\tilde{\nu}H+\ldots$.
This terms comes from a 1-loop triangle diagram (see Fig. \ref{fig1}) involving
$\tilde{\nu}$, $\tilde{\nu}_{R}$ and H \cite{Chacko:2003tf}
\cite{Choi:2003fk} \cite{Antoniadis:1998sd, Delgado:1998qr}.\\

From the diagram, Fig.~\ref{fig1}, we
see that the only possible meaningful contributions can come from
the gaugino zero mode and the right handed sneutrino zero mode,
since $\tilde{\nu}$ and H live in the UV brane. Concentrating on
the dominant wino contribution, we get
\begin{equation}
A_{4}\lambda_{4}=4\lambda_{4}g_{4}^{2}C_{2}(N)\int\frac{d^{4}p}{(2\pi)^{4}}
\frac{1}{p^{2}}\frac{m_{\tilde{W}}}{p^{2}+m_{\tilde{W}}^{2}}
\end{equation}
where $g_{4}^{2}=g_{5}^{2}/R$. Integrating up to the
compactification scale $k e^{-k\pi R}$ (the scale at which SUSY
breaking is transmitted) we get for $A_{4}$ ,

\begin{equation}
\label{eq:A4}
A_{4} \simeq 3\alpha_{2}\frac{m_{\tilde{W}}\log{(ke^{-k\pi R}/m_{\tilde{W}})}}
{2\pi}   .
\end{equation}
where $\alpha_{2}=g_{4,2}/(4\pi)\approx 0.03$. Although the proper
result for $A_4$ can only be obtained after resummation, due to the
presence of the weak gauge coupling and the relatively low
scale for $k e^{-k \pi R}$, the above result provides a very good
approximation for the Wino contribution to $A_4$.
Also, since the beta function of
$\alpha_{2}$ in the MSSM is small, the value of $\alpha_2$ may be
approximately replaced by its weak scale value. \\

  The mass spectrum of squarks and sleptons can be obtained through
gaugino mediation as is done in \cite{Chacko:2003tf} since the
F-term radion contributions to these masses will be smaller.

\section{Results}

Now we have all the necessary ingredients to do soft leptogenesis
in warped extra dimensions. We will constraint our model by
obtaining reasonable values for the lepton asymmetry
$\epsilon_{L}$, the left handed neutrino mass through the see saw
mechanism $m_{\nu}$, requiring that the necessary conditions for
leptogenesis are satisfied
and that the lepton asymmetry is not erased because of the KK modes.\\

As was mentioned in the introduction, we will consider a single
generation of right handed neutrinos, since the effect survives
even in this limiting case. We will drop all flavor indices. Under
these conditions, a CP-violating phase is still present. We can see
that in the following way. Let us write the important terms for CP
violation in the 4D Lagrangian,
\begin{eqnarray}
& &
-\Delta\mathcal{L}_{4D}=\ldots+\lambda_{4}(\nu\nu_{R}H+\nu_{R}h\tilde{\nu}+h\nu\tilde{\nu}_{R})+
g_{4,2}\sqrt{2}(\tilde{\nu}^{*}\tilde{W}\nu+H^{*}\tilde{W}h)+\nonumber\\
& & \frac{1}{2}\tilde{B}_4
\tilde{\nu}_{R}\tilde{\nu}_{R}+\frac{1}{2}(M_{4,IR}+M_{4,UV})\nu_{R}\nu_{R}+
\frac{1}{2}m_{\tilde{W}}\tilde{W}\tilde{W}+h.c.\label{cp}
\end{eqnarray}
where $\tilde{B}_4 = B_4 M_{4} = k\pi F_{T}M_{4,IR}$, and
$m_{\tilde{W}} \propto F_{T}$. The conformal sector of the action
is invariant under an R-symmetry $U(1)_{R}$ and a
Peccei-Quinn-type symmetry $U(1)_{Q}$. In analogy with what was
discussed in Ref.~\cite{Pilaftsis:1999qt}, under $U(1)_{R}$ and
$U(1)_{Q}$ the fields have the corresponding charges listed in
Table~\ref{tab1}.
\begin{table}[h]
\centering
\begin{tabular}{|c||c||c|}
\hline
Field & R-Charge & PQ-Charge\\
\hline \hline
H & 0 & -2\\
h & -1 & -2\\
$\tilde{\nu}$ & 1 & 0\\
$\nu$  & 0 & 0\\
$\tilde{\nu}_{R}$ & 1 & 2\\
$\nu_{R}$ & 0 & 2\\
$\tilde{W}$ & 1 & 0\\
\hline
\end{tabular}
\caption{R-Charges}\label{tab1}
\end{table}
 Then if, for instance, we start with a single phase in the gaugino mass
 $m_{\tilde{W}}$ and assume that the other parameters are real,
since the first line of Eq.~(\ref{cp}) and the Majorana masses for
$\nu_{R}$ are invariant
 under the R-symmetry, we can rotate away the phase in the gaugino
mass parameter by doing a $U(1)_R$
transformation, while  generating a phase on $\tilde{B}_{4}$.
We can also remove the phase in $\tilde{B}_4$ by means of
a $U(1)_{Q}$ superfield rotation of the right handed neutrino
and the Higgs, but we generate a phase in the total Majorana mass
$M_{4}=M_{4,IR}+M_{4,UV}$. So we see that there is no possible way to
eliminate this CP-violating phase, which is therefore physical. In
general, we can identify this phase with
\begin{equation}
\phi = {\rm arg} \left(M_{4} \; m_{\tilde{W}}\tilde{B}_{4}^{*} \right).
\end{equation}
Using the fact that $B_4 \propto m_{\tilde{W}} M_{4,IR}$,
one can easily see that $\phi = {\rm arg}(M_{4} M_{4,IR}^*)$.
We will work in a basis in which the total Majorana mass
$M_4$ and the bilinear mass term $\tilde{B}_4$ are real ($B_4$ real), and
therefore  $\phi$ may be identified with the CP-violating phase
associated with the gaugino mass term. From Eq.~(\ref{eq:A4}) it
follows that the phase $\phi$ is transferred
to the trilinear term $A_4$ at the loop level.
Following \cite{D'Ambrosio:2003wy},  in the limit
$|\lambda_{4}|^{2}A_{4}/4\pi\ll B_{4}/2$ the lepton asymmetry is
approximately given by
\begin{eqnarray}
& &
\epsilon_{L} \simeq
\frac{4\Gamma B_{4}}{4B_{4}^{2}+\Gamma^{2}}\frac{ImA_{4}}{M_{4}}
\Delta_{BF}\\
& &
\Delta_{BF}=\frac{c_{B}-c_{F}}{c_{F}+c_{B}}
\label{epsilonL}
\end{eqnarray}
where $c_{B}$ and $c_{F}$ represent the fermionic and bosonic decay
channel rates with
final states $f=h\nu$ and $f=H\tilde{\nu}$ respectively,
and $\Delta_{BF}\approx 0.8$ for $T=1.2M_{4}$.\\

Here an important point must be raised. What we measure from
experiments is the ratio
\begin{equation}
\frac{n_{B}}{n_{\gamma}}\approx 7 \frac{n_{B}}{s}\approx 6 \times 10^{-10}
\end{equation}
where $n_{B}$ is the baryon number density, $n_{\gamma}$ the
photon number density and s is the entropy density (for a more
detailed discussion see section~\ref{sec:numres}). After
production, and after the action of weak sphaleron effects,
the total baryon asymmetry is fixed and therefore the
baryon to entropy ratio remain constant. In that sense, the
present measurement of $n_B/s$ reflects the primordially generated
value.\\

As shown in \cite{Grossman:2005yi}, we can write the baryon to entropy ratio as
\begin{equation}
\frac{n_{B}}{s}=-\left(\frac{24+4n_{H}}{66+13n_{H}}\right)Y^{eq}_{\tilde{\nu}_{R}}\xi\left[\frac{4\Gamma|B_{4}|}{4|B_{4}|^{2}+
\Gamma^{2}}\right]\frac{|A_{4}|}{M_{4}}\sin (\phi)\label{baentro}
\end{equation}
where the first factor takes into account the reprocessing of the
B-L asymmetry by sphaleron transitions, $n_{H}$ is the number of
Higgs doublets which is equal to 2,
$Y^{eq}_{\tilde{\nu}_{R}}=45\zeta (3)/(\pi^{4}g_{*})$ where
$g_{*}$ is the number of thermalized degrees of freedom, $\xi$ is
an efficiency parameter that slightly depends on the production
mechanism for the right handed neutrinos and $\phi$ is the CP
violating phase defined above ($\sin (\phi)\simeq 1$). Assuming
thermal production, $\xi$ is suppressed for small and large
$m_{\nu}$ because of insufficient $\nu_{R}$ production and strong
washout effect, respectively. The maximum value is
$\mathcal{O}(0.1)$
for $m_{\nu}\simeq 10^{-(3-4)}$ eV.\\

Now, as seen from Eq.~(\ref{baentro}), during the radiation dominated era the entropy density $s$ is given by
\begin{equation}
s\sim g_{*} T^{3}
\end{equation}
 So we see from here that the towers of KK modes, if thermalized, will
contribute to $g_{*}$. Since the mechanism of leptogenesis depends
on the decay of the chiral right-handed (s)neutrino zero mode, the
presence of the KK towers will not induce an extra contribution to
the lepton asymmetry. Therefore, the effect of the KK modes will
be in general to dilute the leptonic asymmetry. Such a dilution,
if large, would make it difficult to obtain the experimentally
observed baryon asymmetry. Therefore, for simplicity, we shall
assume $k e^{-k \pi R} \simgt M_4$ and therefore only the zero
mode will be in thermal equilibrium at temperatures of the order
of the Majorana neutrino mass, $T\sim M_4$. Values of $kR \sim 10$
satisfying  $k e^{-k\pi R} \simeq M_4$,  are also consistent
with the ones necessary to stabilize the vacuum expectation value
of $T(x)$~\cite{Goldberger:1999uk}, and therefore from now on we
shall assume that the latter is
satisfied.\\

Under the above conditions, the light neutrino properties will be governed
by the right-handed neutrino mass. Namely,
through the implementation of the see-saw mechanism in warped extra
dimensions~\cite{Huber:2003sf},
the left handed neutrino mass is given by
\begin{equation}
m_{\nu}\sim \frac{v^{2}|\lambda_{4}|^{2}}{M_{4}}
\end{equation}
where $v=\langle H(x)\rangle \simeq 174$ GeV  is the expectation value of Higgs field, and we are assuming $\tan \beta \gg 1$.\\

To satisfy the out of equilibrium condition  in the decay of the
right handed sneutrino, we should have a decay rate,
$\Gamma=M_{4}|\lambda_{4}|^{2}/4\pi$ that is not much faster than
the expansion rate of the universe H. Since at $T \simeq M_4$ only
the zero modes are in thermal equilibrium, we can essentially use
 four dimensional cosmology, ignoring the extra dimensional contributions
of the gauge and neutrino fields propagating in the bulk.
Therefore, we shall use the conventional four dimensional
expression $H=1.66g_{*}^{1/2}T^{2}/M_{P}$ at the time when $T\sim
M_{4}$ ($M_{P}=4.2\times 10^{18}$ GeV and  $g_{*}$ counts the
number of
degrees of freedom (d.o.f) in thermal equilibrium).\\

We can improve the above approximation by writing
an expression for the number of thermalized degrees of freedom
as $g_{*}= N_{KK}\times g_{*,1}+g_{*,2}$, where $N_{KK}$ counts the
number of excited KK levels that are in thermal equilibrium.\\

 The masses of the first KK excited levels of a vector superfield
(the major contribution to the KK states comes from the gauge modes),
in the limit $KR\gg 1$ and $m_{n}\ll k$, are given by
\begin{equation}
m_{n,V}\simeq(n-\frac{1}{4})\pi ke^{-k\pi R}\label{mkk}
\end{equation}
 where $n=1,2,\ldots$ is the KK level.\\

In the UV brane, there are 45 chiral superfields (quarks and leptons), 2 Higgs
doublets which include 4 chiral superfields. Therefore, there are 49 chiral
superfields, each one containing 4 physical degrees of freedom (2 fermionic
and 2 bosonic). The total number of effective degrees of freedom
adds to $g_* = 98 \times (1 + 7/8) \simeq 184$. The zero mode fields
of the right handed neutrino and  gauge superfields contribute in the
following
way. There are 12 gauge fields (8 from QCD and 4 from electroweak), each one
having two polarizations since they are massless. Therefore, the gauge
superfields
zero-modes contribute $12 \times2 \times 15/8 =45$
effective degrees of freedom, where the last factor of~15/8
comes from SUSY. The right handed neutrino zero modes superfields belong to
3 families and
they are Weyl fermions (2 degrees of freedom). Thus, they contribute as
$3\times2\times 15/8 \simeq 11$
degrees of freedom. So we conclude that $g_{*,2}=240$.\\

 In the case of the KK towers we have to remember that fields then are part of
$\mathcal{N}=2$ SUSY.
Therefore each tower (counting gauge and three right-handed neutrino
superfields which are the only ones that contribute to them)
will add $60\times 15/8 \simeq 112$
degrees of freedom and thus  $g_{*,1}=112$. To calculate $N_{KK}$ we go through
the following derivation. The entropy density $s$ is given by the expression
\begin{equation}
s=\frac{\rho+p}{T}
\end{equation}
Now $\rho$ and $p$ can be written as
\begin{eqnarray}
& &
\rho=\frac{g}{2\pi^{2}}\int_{m}^{\infty}\frac{(E^{2}-m^{2})^{1/2}}{e^{(E-\mu)/T}-1}E^{2}dE\label{rho}\\
& &
p=\frac{g}{6\pi^{2}}\int_{m}^{\infty}\frac{(E^{2}-m^{2})^{3/2}}{e^{(E-\mu)/T}-1}dE\label{p}
\end{eqnarray}
 where we are doing the calculations for bosons~\footnote{the results are basically the same for fermions}. We will assume that $T\gg \mu$. In the relativistic limit, $T\gg m$, $\rho=(\pi^{2}/30)gT^{4}$ and $p=\rho/3$. We define $N_{KK}|_{n}$ as%
\begin{equation}
N_{KK}|_{n}=\frac{s|_{n,non-rel}}{s|_{n,rel}}
\end{equation}
for each KK level parameterized by $n$ in Eq.~(\ref{mkk}),
$s|_{n,rel} = g (2 \pi^2/45) T^3$ is the entropy contribution in
the relativistic limit, and we use the full expressions,
Eqs.~(\ref{rho})-(\ref{p}) to calculate the non-relativistic
entropy contribution, $s|_{non-rel}$. Therefore $N_{KK}$ is given
by
\begin{equation}
N_{KK}=\sum_{n=1}^{+\infty}N_{KK}|_{n}
\end{equation}
For values of $m \gg T$, the effective number of degrees of
freedom associated with a given specie  is suppressed by a factor
$(m/T)^{3/2} \exp(-m/T)$.\\

Using the above expression for the KK mode masses, with
$m_{1,V} \simeq 2.3 k \exp(-k \pi R)$,
we can easily perform the sum. It is straightforward to prove that, provided
$T < 2.3 k \exp(-k \pi R)$ the value of $N_{KK}$ will remain lower
than one, leading to only a small modification of the effective
number of degrees of freedom at the freezing temperature.
We shall require that this relation is fulfilled, in order to avoid a dilution
of the baryon asymmetry. In addition, since the freezing temperature
$T \simeq M_4$, the above relation ensures that the zero right-handed
neutrino mode will have only small mixing with the heavier KK modes,
increasing the validity  the approximations used in this work to
generate the Majorana mass contribution to the right-handed neutrino
zero mode.\\

The out of equilibrium condition then reads,
\begin{equation}
M_{4}/|\lambda_{4}|^{2}\gtrsim
\frac{M_{P}}{4\pi\times1.66\times(1.2)^{2}\sqrt{g_{*}}}
{\rm GeV}\label{out}
\end{equation}
On the other hand, the sneutrino decay should occur before the
electroweak phase transition, when sphalerons, responsible for the
conversion of lepton asymmetry into baryon asymmetry are still
active, $\Gamma>H(T\sim 100{\rm GeV})$
\begin{equation}
M_{4}|\lambda_{4}|^{2}\gtrsim 3\times10^{-13}{\rm GeV}
\end{equation}
\subsection{Analytical Estimates}

Taking into account the above constraints, we can obtain information about
the fundamental parameters of the theory from an analytical point of view.
Assuming that the right handed neutrino is located away from the infrared
brane,
\begin{equation}
e^{1/2+c_{\nu_{R}}}\gg1\label{cond1}
\end{equation}
and that the five dimensional Yukawa coupling $\lambda_5 \simeq a/\sqrt{k}$,
with $a$ a number of order one, we obtain,
\begin{eqnarray}
& &
\lambda_{4}=a\sqrt{1+2c_{\nu_{R}}}e^{-(1/2+c_{\nu_{R}})k\pi R}\\
& &
M_{4,UV}=\frac{\lambda_{4}^{2}}{a^{2}}kRM_{1}\simeq M_4 \\
\end{eqnarray}

We can now see that the out of equilibrium condition, Eq.~(\ref{out}),
sets the scale for $M_{1}$
\begin{equation}
M_4/\lambda_{4}^{2} \simeq \frac{kRM_{1}}{a^{2}}.
\label{rel}
\end{equation}
Hence, as emphasized in the introduction, this condition is naturally
satisfied for values of
$M_{1}\sim 10^{14}-10^{15}$ GeV close to the GUT scale.
At the same time, the same parameter $M_1$ sets the scale of the neutrino
mass parameter. Indeed, since
\begin{equation}
M_4/\lambda_4^2 =\frac{v^{2}}{m_{\nu}}
\label{neuM1}
\end{equation}
and therefore, for the above values of $M_1$, the neutrino mass
parameter acquires phenomenologically acceptable values
$m_{\nu}\sim 10^{-3}$ eV.\\

In order to determine the value of the remaining parameters, we should
take into account the conditions necessary for the realization of the
soft leptogenesis scenario. As it is clear from Eq.~(\ref{epsilonL}),
these depend on the specific values of the
bilinear and trilinear parameters derived before, as well as the
decay width. The relevant parameters are given by
\begin{eqnarray}
& &
M_{4,IR} \simeq
\frac{\lambda_{4}^{2}}{a^{2}}kRM_{2}e^{2c_{\nu_{R}}k\pi R}\\
& &
B_{4}=2k\eta e^{-k\pi R}\frac{M_{4,IR}}{M_{4,UV}}=2k\eta e^{(2c_{\nu_{R}}-1)k\pi R}\frac{M_{2}}{M_{1}}\\
& &
m_{\tilde{W}} \simeq \frac{\eta k}{\pi kR}e^{-k\pi R}\\
& &
\Gamma=\frac{M_{4,UV}\lambda_{4}^{2}}{4\pi}=
\frac{M_{1}kRa^{2}(1+2c_{\nu_{R}})^{2}e^{-(2+4c_{\nu_{R}})k\pi R}}{4\pi}
\end{eqnarray}
The primordial lepton asymmetry is maximized when the decay with is of
the order of $B_4$. Requiring the parameters to be close to the
resonance condition for $\epsilon_{L}$ ($\Gamma=2B_{4}$)
we obtain the following relationship
\begin{equation}
\frac{M_{2}k}{M_{1}^{2}a^{2}}=\frac{kR}{16\pi\eta}(1+2c_{\nu_{R}})^{2}
e^{-(1+6c_{\nu_{R}})k\pi R}\label{resonance}
\end{equation}
Furthermore, now asking that at the specific temperature of soft
leptogenesis there are very few KK excited states, $T\sim
M_{4}\sim ke^{-k\pi R}$, implies
\begin{equation}
\frac{k}{M_{1}}\sim(1+2c_{\nu_{R}})kRe^{-2c_{\nu_{R}}k\pi R}\label{fewkk}
\end{equation}
Combining Eq.~(\ref{resonance}) with Eq.~(\ref{fewkk}) we arrive at
the following relation
\begin{equation}
M_{2}\sim \frac{ka^{2}}{16\pi \eta kR}e^{-(1+2c_{\nu_{R}})k\pi R}\label{m2}
\end{equation}
A further constraint we need to impose on the model is that the mass
of the NLSP which is the stau $\tilde{\tau}_{1}$, as is the case in
\cite{Chacko:2003tf}, be in accordance with experimental constraints.
We use the RGE at one loop
\cite{Castano:1993ri},
\begin{equation}
\frac{dm_{\tilde{\tau_{1}}}^{2}}{dt}=\ldots+\frac{1}{8\pi^{2}}
\left(-\frac{12}{5}g_{4,1}^{2}m_{\tilde{B}}^{2}\right)+\ldots
\end{equation}
where $g_{4,1}$ is the $U(1)_{Y}$ 4D hypercharge, $m_{\tilde{B}}$
is the bino mass, $t=ke^{-k\pi R}/m_{\tilde{\tau}}$, and we only
included the relevant term. To avoid experimental constraints,
$m_{\tilde{\tau}_{1}}\simgt 100$ GeV or bigger, we need to have a
gaugino mass $m_{\lambda_{1}}\simgt 500$ GeV.\\

Taking the compactification scale $ke^{-k\pi R}\sim M_{4}$ but at
the same time having a gaugino mass of $\mathcal{O}$(500) GeV and
furthermore having a small gravitino mass ($m_{3/2}<16$ eV from
cosmological constraints, see next section) fixes the values of
$kR$ and $\eta$. With this requirements,  we find that  the value
of the trilinear term $A_{4}\simeq \mathcal{O}(60)$ GeV. Now, in
the case of resonance, $\Gamma \simeq 2 |B_4|$, this implies a
maximum value for $M_{4}$ since $\epsilon_{L}\simeq
A_{4}/M_{4}\simeq 10^{- 6}$. But from Eq.~(\ref{neuM1}) and the
discussion about the efficiency parameter $\xi$ following
Eq.~(\ref{baentro}), we see that by fixing $m_{\nu}\simeq
10^{-(3-4)}$eV we completely determine the value of $\lambda_{4}$.
Thus in turn, from Eq.~(\ref{rel}), draws us to fully fix the
ratio $M_{1}/a^{2}$. Since $a\simeq \mathcal{O}$(1), we see that
$M_{1}\simeq \mathcal{O}(10^{(14-15)})$ GeV, of the order of the
unification scale, as stated in the introduction and previously on
this section. Rewriting Eq.~(\ref{m2}) as
\begin{equation}
M_{2}=\frac{k\lambda_{4}^{2}}{16\pi\eta kR(1+2c_{\nu_{R}})}
\end{equation}
we see that $M_{2}$ is also almost fully fixed by the choice of
$m_{\nu}$ except for a mild dependence on $1/(1+2c_{\nu_{R}})$.
Thus, we have shown that given the conditions described above, all
parameters are fixed by choosing values for $m_{\nu}$
and the parameter $a$.\\

\subsection{Numerical Results}
\label{sec:numres}

The numerical ratio of the baryon density to the entropy density
may be obtained experimentally by two methods. Firstly, by the
requirement of consistency between the observed and the predicted
abundance of primordial elements by the Standard Big-Bang
Nucleosynthesis model~\cite{Olive:1999ij}. This  leads to a value
of the baryon to photon ratio~\cite{Fields:2006ga},
\begin{equation}
4.5 \times 10^{-10} \simlt \frac{n_B}{n_{\gamma}} \simlt 6.5 \times 10^{-10}.
\end{equation}
The second method is related  to the baryon energy density determination by the
WMAP experiment~\cite{Spergel:2006hy},
\begin{equation}
\label{eq:wmap}
\Omega_b h^2 = (2.233 \pm 0.072) 10^{-2}.
\end{equation}
~\\
Considering the relation between the entropy and the photon density,
$s \simeq 7 n_{\gamma}$~\footnote{Although the addition of the gravitino
increases the total entropy density, this increase is very small due
to the large dilution factor associated with the gravitino decoupling
temperature, $T_D > 1 GeV$.},
the Big-Bang Nucleosynthesis results translate into a value of
\begin{equation}
\label{eq:broader}
6.5 \times 10^{-11} \simlt n_B/s \simlt 9.5\times 10^{-11} ,
\end{equation}
with a narrower band of values, around $9 \times 10^{-11}$ being
selected if only the WMAP values are considered. The WMAP result,
Eq.~(\ref{eq:wmap}), may be slightly modified (up to values of
$\Omega_b h^2 \simeq 0.019$) by assuming different shapes of the
power spectrum~\cite{Fields:2006ga, Blanchard:2003du}. We shall
require that the baryon number to entropy density ratio that we
compute is within the broader range given above, (\ref{eq:broader}).
However, values within the WMAP allowed band, Eq.~(\ref{eq:wmap}),
may always be obtained by appropriate tuning of the parameters of
the model.\\

 In  Tables~\ref{tab2} and~\ref{tab3} we give the
results from numerical computations for two different acceptable
points in parameter space. The input parameters are listed on the
left column and the output on the right column.
\begin{table}[h]
\centering
\begin{tabular}{|c||c|}
\hline
Input 1 & Output 1 \\
\hline \hline
$c_{\nu_{R}}=-0.12$  & $\lambda_{4}=1.98\times 10^{-5}$ \\
$kR=8$ & $ke^{-k\pi R}=1.216\times 10^{7}$ GeV \\
$M_{1}=3\times10^{14}$ GeV & $M_{4,UV}=9.23\times10^{6}$ GeV\\
$M_{2}=1\times 10^{10}$ GeV & $M_{4,IR}=0.73$ GeV\\
$k=1\times10^{18}$ GeV & $m_{\lambda_{1}}=484$ GeV \\
$\lambda=0.32/\sqrt{k}$ & $A_{4}=70.11$ GeV \\
$\eta=10^{-3}$ & $m_{\nu}= 1.29\times10^{-3}$ eV\\
& $B_{4}=0.0019$ GeV \\
& $\Gamma_{4}=0.00029$ GeV \\
& $\epsilon_{L}=1.12\times10^{-6}$\\
& $m_{3/2}\approx20$ eV\\
& $M_{4}/\lambda_{4}^{2}=2.44\times 10^{16}$ GeV\\
& $N_{KK}=0.55$\\
& $n_{B}/s\simeq 7.2\times 10^{-11}$\\
\hline
\end{tabular}
\caption{Results}\label{tab2}
\end{table}
\begin{table}[h]
\centering
\begin{tabular}{|c||c|}
\hline
Input 2 & Output 2\\
\hline \hline
 $ c_{\nu_{R}}=-0.105$ &  $\lambda_{4}=1.54\times 10^{-5}$ \\
 $kR=8.42$ & $ke^{-k\pi R}=9.75\times 10^{6}$ GeV\\
 $M_{1}=1.1\times10^{15}$ GeV & $M_{4,UV}=6.148\times10^{6}$ GeV\\
 $M_{2}=10^{10}$ GeV & $M_{4,IR}=0.21$ GeV\\
 $k=3\times10^{18}$ GeV & $m_{\lambda_{1}}=479$ GeV \\
 $\lambda=0.6/\sqrt{k}$ & $A_{4}=66$ GeV\\
 $\eta= 1.3\times10^{-3}$ & $m_{\nu}= 1.176\times 10^{-3}$ eV\\
 & $B_{4}=0.00089$ GeV\\
 & $\Gamma_{4}=0.00011$ GeV\\
 & $\epsilon_{L}=1.42\times10^{-6}$ \\
 & $m_{3/2}\approx17$ eV\\
 & $M_{4}/\lambda_{4}^{2}=3.0825\times 10^{16}$ GeV\\
 & $N_{KK}=0.40$\\
 & $n_{B}/s\simeq 9.61\times 10^{-11}$\\
\hline
\end{tabular}
\caption{Results}\label{tab3}
\end{table}

From the tables we see that, as emphasized before,
$M_{4}=M_{4,UV}+M_{4,IR}\approx M_{4,UV}$. We also notice that as
we lower $M_{4}$ we don't need to be so close to the resonance
condition which, as said before, is fulfilled when $\Gamma=
2|B_{4}|$. Moreover, $M_{4}/\lambda_{4}^{2}$ satisfies the out of
equilibrium condition, Eq.~(\ref{out}).  All necessary conditions
for soft leptogenesis are satisfied and we get a left-handed
neutrino mass which is of the order of the one associated with the
values
of the eigenstate mass differences implied by solar neutrino experiments.\\

To calculate the gravitino mass we used the approximate formula
\begin{equation}
m_{3/2}\approx \frac{\eta k^{2}e^{-2k\pi R}}{\sqrt{3}M_{P}}
\end{equation}
and we have required that the gravitino mass obtained from this expression is
lower than about 20~eV, to be consistent with  astrophysical and cosmological
bounds (see below).\\

 In Tables~\ref{tab2},~\ref{tab3}, we have chosen a value of
$k \simeq 10^{18}$~GeV,
of the order of the fundamental Planck scale. As the value of $k$~is lowered, we see that $N_{KK}\simgt 1$ and $M_{2}\simlt 10^{10}$ GeV. We provide an example, with values of
$k \simeq 2\times 10^{17}$~GeV
in table ~\ref{tab4}.\\
\begin{table}[h]
\centering
\begin{tabular}{|c||c|}
\hline
Input 3 & Output 3\\
\hline \hline
 $ c_{\nu_{R}}=-0.105$ &  $\lambda_{4}=2.138\times 10^{-5}$ \\
 $kR=7.6$ & $ke^{-k\pi R}=8.546\times 10^{6}$ GeV\\
 $M_{1}=3.3\times10^{14}$ GeV & $M_{4,UV}=1.27\times10^{7}$ GeV\\
 $M_{2}=3\times10^{9}$ GeV & $M_{4,IR}=0.77$ GeV\\
 $k=2\times10^{17}$ GeV & $m_{\lambda_{1}}=537$ GeV \\
 $\lambda=0.3/\sqrt{k}$ & $A_{4}=74.4$ GeV\\
 $\eta=1.5\times 10^{-3}$ & $m_{\nu}= 1.086\times 10^{-3}$ eV\\
 & $B_{4}=0.0015$ GeV\\
 & $\Gamma_{4}=0.00046$ GeV\\
 & $\epsilon_{L}=1.71\times10^{-6}$ \\
 & $m_{3/2}\approx15$ eV\\
 & $M_{4}/\lambda_{4}^{2}=2.786\times 10^{16}$ GeV\\
 & $N_{KK}=1.38$\\
 & $n_{B}/s\simeq 8.35\times 10^{-11}$\\
\hline
\end{tabular}
\caption{Results}\label{tab4}
\end{table}

In models of low energy supersymmetry breaking, like the one under
consideration, the gravitino is the lightest SUSY particle and
thus it is stable. Therefore gravitinos generated at high
temperature contribute to the matter density of the universe. For
masses $m_{3/2}\lesssim 100$~eV the goldstino component of the
gravitino has large interaction with the MSSM particles, and
therefore the gravitino can thermalize at high temperature. The
number density in this case is just the equilibrium value and,
taking into account the diluting effect of the decoupling of heavy
particles,  the energy density is approximately given by
$\Omega_{3/2}h^{2}\sim 0.1(m_{3/2}/100\;eV)$, satisfying
cosmological bounds. Since these gravitinos are warm, from
Lyman-$\alpha$ forest and WMAP data in order for them not to smear
out the density perturbations on the matter power spectrum at
small scales their masses are excluded from the region
16~eV$\lesssim m_{3/2}\lesssim$~100~eV. The gravitino mass in our
scenario falls naturally in the 1--100~eV range, and satisfies
these constraints for a broad range of parameters, as
shown by the specific examples above.\\

In the above, we have considered a model of gaugino mediation in
warped extra dimensions, in which the matter fields are localized
on the UV brane, while the dominant supersymmetry breaking
contribution is localized on the IR brane. A question arises about
the possible origin of the supersymmetry Higgsino mass term~$\mu$
within such scenario. Since the value of the gravitino mass is
much lower than in supergravity mediated scenarios, the
Giudice-Masiero mechanism~\cite{Giudice:1988yz} won't provide a
sufficiently large mass. A logical possibility is the addition of
a singlet in the spectrum, which couples to the Higgs superfields
and induces a $\mu$-term by acquiring a v.e.v. If this singlet is
localized, it will acquire a negative supersymmetry breaking
squared mass term by radiative corrections. Since supersymmetry is
mediated by gaugino interactions, this is a higher-loop effect,
and numerically the mass values are too small to lead to a
phenomenologically acceptable  $\mu$ parameter, for natural values
of the Higgs and singlet couplings. However, as has been
previously done in similar low energy supersymmetry breaking
models~\cite{Nilles:1997me}, one can make use of supergravity
induced tadpole contributions and the compensator field
(\ref{eq:compensator}), whose F-term is $F_{\phi}\simeq m_{3/2}$,
to lead to an acceptable value of $\mu$. As an alternative to the
localized singlet field, one can also consider the case of a
singlet field propagating in the bulk of the warp extra dimension.
Although in this case the result depends on the precise
localization of the singlet zero mode in the bulk, an acceptable
$\mu$-term may be obtained for reasonable values of the bulk mass
parameters.

\section{Conclusions}

In this article, we have studied the possibility of realizing the mechanism of
soft leptogenesis within the context of warped extra dimensions. We have
assumed that all the  quark and lepton fields  are localized on the UV brane,
while the gauge fields and the right-handed neutrinos propagate into the
extra dimension. Assuming the presence of localized Majorana mass terms
on the UV and IR branes, we have shown that the condition of out of equilibrium
may be naturally fulfilled by assuming that the UV Majorana mass term is of
the order of the GUT scale. Furthermore, for the same conditions, the neutrino
mass acquires phenomenologically acceptable values, and of the order of
the ones necessary to maximize the baryon asymmetry result.\\

Soft supersymmetry breaking parameters for the gauginos and for the
right-handed sneutrinos are generated by the auxiliary
component of the radion field, which acquires a non-vanishing vacuum
expectation
value localized on the IR brane. Loop effects are responsible for the
generation of supersymmetry breaking parameters for the rest of the quark,
lepton  and Higgs superfields. Although the right-handed stau
becomes the lightest
standard model superpartner, the lightest superparticle is given by the
gravitino
which becomes naturally light within this framework.
Then, the collider phenomenology
becomes similar to the one of gauge-mediated models with a light stau NLSP \cite{Giudice:1998bp}.\\

We have shown that, provided a relative phase exist between the two localized
Majorana masses, the physical CP-violating phase necessary for the
realization of soft
leptogenesis is generated. An effective Majorana mass $M_4$ smaller than
about $10^8$~GeV, as necessary for the realization of this scenario,
is naturally generated by a proper localization of the right-handed
neutrino zero modes. Moreover, this localization is also effective
in avoiding the
dilution of the baryon asymmetry by the entropy generated by the KK towers
provided the Majorana mass is smaller than the
local curvature term in the IR brane, $M_4 \simlt k \exp(-k\pi R)$.
Finally, the condition that the gaugino masses are at the TeV scale fixes
the size of $F_T$. The resulting gravitino mass is of the order of a
few~eV, which satisfy the relic density and long range structure
constraints dictated by cosmology.\\

A proper baryon asymmetry may be generated under the above
conditions, provided the effective bilinear term $B_4$ is of
the order of the sneutrino decay width. Due to the smallness
of the Yukawa couplings, this implies a value of $B_4$ smaller
than about 100~MeV. In our model the value of $B_4$ is determined
by the ratio of the localized Majorana mass term, and successful
leptogenesis is achieved for values of the localized UV and IR
mass terms $M_1 \simeq 10^{15}$~GeV and
$M_2 \simeq 10^{10}$~GeV, respectively.

\vspace{1.5cm}
~\\
\large{\textbf{Acknowledgements:}} \normalsize
We wish to thank Tim Tait and Eduardo Ponton
for helpful discussions.
Work at ANL is supported in part by the US DOE, Div.\ of HEP, Contract
W-31-109-ENG-38.

\appendix

\bibliographystyle{prsty}
\bibliography{lepto}

\end{document}